\documentclass[epj,twocolumn]{webofc}
\usepackage[varg]{txfonts}   % Web of Conferences font
\usepackage{widetext}
\usepackage[export]{adjustbox}
\woctitle{Powders \& Grains 2017}
\begin{document}
\title{Thermal properties of an impurity immersed in a granular gas of rough hard spheres}

\author{\firstname{Francisco} \lastname{Vega Reyes}\inst{1,2}\fnsep\thanks{\email{fvega@unex.es}} \and
        \firstname{Antonio} \lastname{Lasanta}\inst{1,3} \and
        \firstname{Andr\'es} \lastname{Santos}\inst{1,2} \and
        \firstname{Vicente} \lastname{Garz\'o}\inst{1,2}
}

\institute{Departamento de F\'isica, Universidad de Extremadura, 06071 Badajoz, Spain 
\and
Instituto de Computaci\'on Cient\'ifica Avanzada (ICCAEx), Universidad de Extremadura, 06071 Badajoz, Spain
\and G. Millán Institute, Fluid Dynamics, Nanoscience and Industrial Mathematics Department of Materials Science and Engineering and Chemical Engineering Universidad Carlos III de Madrid Leganés, Spain}

\abstract{%
  We study in this work the dynamics of a granular impurity immersed in a low-density granular gas of identical particles. For description of the kinetics of the granular gas and the impurity particles we use the rough hard sphere collisional model. We take into account the effects of non-conservation of energy upon particle collision. We find an (approximate) analytical solution of the pertinent kinetic equations for the single-particle velocity distribution functions that reproduces reasonably well the properties of translational/rotational energy non-equipartition. We assess the accuracy of the theoretical solution by comparing with computer simulations. For this, we use two independent computer data sets, from molecular dynamics (MD) and from Direct Simulation Monte Carlo method (DSMC). Both approach well, with different degrees, the kinetic theory within a reasonable range of parameter values. 
}
\maketitle
\section{Introduction}
\label{submit}
The dynamics of matter composed by large sets of macroscopic particles (sizes larger than $\sim 1\mu\mathrm{m}$ \cite{B54}) has always been of interest \cite{F31,R85} since particulate matter is present, under a huge variety of forms and composites, in many human-activity and natural processes. Moreover, the activity of this research field has grown enormously over the last decades \cite{JNB96b, G99, K99, AT06}, becoming an important area of study in the field of soft matter physics \cite{G92,CL00}. In particular, the control of granular dynamics has a large impact on a variety of industries, including some high-technology related interesting studies \cite{QZKUFG15,LUKKG09,DP06}. It is worth to mention also that one of the most ubiquitous problems in granular dynamics related industry is the selection of mechanically peculiar particles in a set of otherwise identical grains; i.e., \textit{granular segregation} \cite{RSPS87}. 

The handling of grains usually requires mechanical input. A variety of mechanisms may be used, of which vibration \cite{OU98} and air-fluidization \cite{LBLG00} are the most common procedures in both fundamental research and granular-matter-related industry. Furthermore, the study of granular matter represents an important challenge for fundamental physics since it has been known from quite some time that soft matter \cite{G92} theories in which the basic constituents are of molecular/atomic nature (and thus particle collisions are elastic) can eventually apply in a generalized form to granular dynamics. Examples range from symmetry breaking processes like the one described in the Kosterlitz-Thouless-Halperin-Nelson-Young (KTHNY) theory \cite{KT73,HN78,Y79} (the so-called hexatic phase \cite{AW62} has also been found in granular systems \cite{OU05}) to kinetic theory of gases or hydrodynamics \cite{D01} at the level of both the associated transport coefficients \cite{BDKS98,GD99,L05} and the hydrodynamic phenomena \cite{VU09}. 

We focus here on the kinetic theory aspects of the granular gas dynamics (a low density and thermalized granular system). It is nowadays widely accepted that a low density granular system can be accurately described by either the Boltzmann equation (low density) or the Enskog equation (moderate density), in its inelastic version \cite{BDS97}. It was P. K. Haff who conceptualized the so-called \textit{granular gas}, defining mathematically its reference state as well, back in 1983 \cite{H83}. Haff defined this state as a system of macroscopic particles at low density that undergo inelastic binary collisions. The system is isolated and its initial state is spatially uniform, this uniformity being preserved in time. This produces a continuously decaying uniform temperature. The cooling rate of the system, as Haff showed \cite{H83}, depends on the degree of inelasticity in the collisions. This is nowadays known as \textit{Haff's law} \cite{SG98}, and his reference state for the granular gas is known as the \textit{homogeneous cooling state} (HCS). The HCS has become the theoretical basis for the development of the kinetic theory of granular gases. There is actually a rigorous experimental proof of the existence of the HCS and the accuracy of Haff's law \cite{MIMA08}.

As we know, the kinetic theory of non-uniform gases accounts for the transport phenomena of non-equilibrium states \cite{CC70}. Starting out of the work by J. C. Maxwell for a gas under equilibrium state, this theory was laid by D. Hilbert \cite{H12}, who set several key mathematical concepts (notably, that of the \textit{normal solution}, to which the non-uniform gas tends after a quick transient, regardless of the initial state). Afterwards, the theory was simultaneously developed and extended by the works of S. Chapman, D. Enskog and D. Burnett \cite{H12,S16,B35}, and finally Ernst and coworkers developed the Enskog theory for denser systems \cite{BE79}. Years later, these theories were extended also to the granular gas \cite{BDKS98,SG98,GD99,L05}, finding consistently the corresponding set of transport coefficients. This, in turn allowed for the deployment of a generalized fluid mechanics, able to additionally describe steady granular flows far from equilibrium. %It has been proven, interestingly, that the hydrodynamics of steady granular flows bears remarkable formal analogy with that of the molecular gases \cite{VU09,VSG10}. Respect to granular hydrodynamics, there is a recent and interesting experiment \cite{GPVP16}, where the authors describe a new type of thermal convection that is induced by two competing perpendicular thermal gradients. What's most interesting is that this thermal convection mechanism is generic, and could be also reproduced in molecular fluids (as ongoing theory work shows \cite{GPVP16_2}), hence yielding another twist in the over-exploited field of B\'enard convection, and variants \cite{B78,B85,K93}. Apparently, and surprisingly, it seems that this mechanism for convection generation has not been used previously in applications nor studied in theory \cite{B78,C81}, hence the potentiality of this recent result.

\begin{figure}
\includegraphics[width=0.35\textwidth,center]{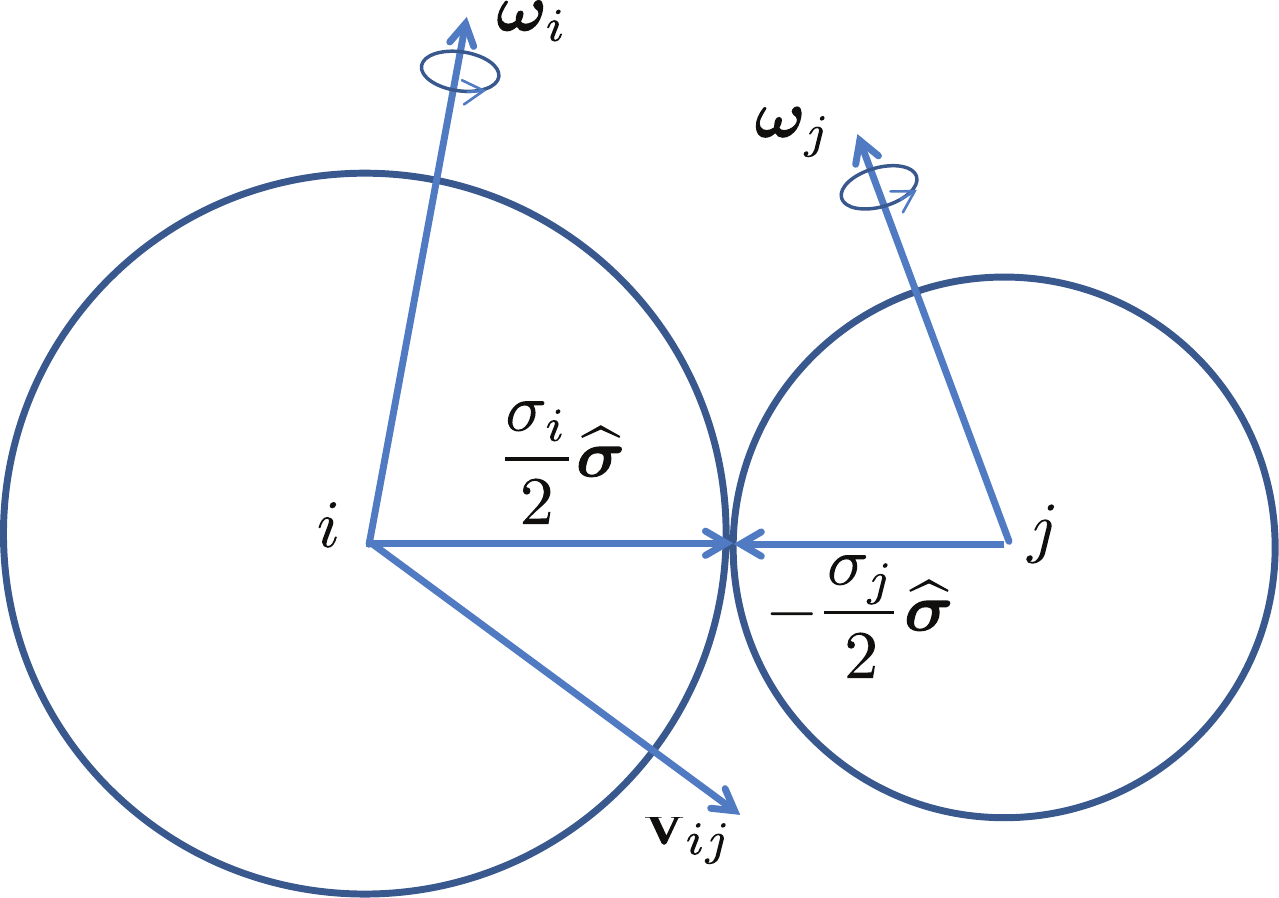}
\caption{The rotational and translational velocities of two grains at collision. Here, $\mathbf{v}_{ij}=\mathbf{v}_i-\mathbf{v}_j$.}
\label{sketch} % Give a unique label
\end{figure}

All of these studies have referred to the simple smooth hard sphere collisional model. However, several important experimental features of granular collisions can only be captured with a more elaborate collisional model \cite{FLCA93}. In particular, particle spin (i.e., the grains angular velocities) can have an important role in the global properties of the granular gas dynamics \cite{AT06}, see Fig.\ \ref{sketch}. A rough hard- sphere collisional model with two velocity-independent coefficients (tangential, $\beta$, and normal, $\alpha$, restitution coefficients) seems to describe more accurately a generic collision between two \text{dry grains} \cite{K10}.

For this reason, and now taking into account the particle rotational degrees of freedom, more recent theoretical studies have revisited homogeneous states like the HCS \cite{BPKZ07,SKG10,VSK14,RA14}, or the steady state produced by a uniform stochastic thermostat \cite{VS15}. In particular, a recent work \cite{SKG10} develops results for binary mixtures. This includes the case where one of the species is present in a very small proportion, as compared to the other species (the so-called tracer limit). This situation (the tracer limit for a binary mixture of inelastic rough hard spheres) is the one we analyze in the present work. The objective of this work is to present a comparison between the existing theory for this system and new simulation data, obtained from Monte Carlo simulation (direct simulation Monte Carlo method, DSMC) of the Boltzmann-Lorentz equation for the granular impurity and also from molecular dynamics (MD) simulations.

%Please submit source files directly to the conference organizers. 
%
%If the conference editors chose to provide print-ready PDF documents to the publisher, you have to submit high-resolution PDF file with all fonts embedded (see PDF guidelines). Remember that no final correction will be made by the publisher. If the organizers chose to use an editorial secretary from the publisher to manage the formatting of the text, please, send the source files to the conference organizers.  
%Send also to the organizers the license agreement signed by at least one author of the article.

\section{Theory and simulation outline}
\label{outline}

Let us consider two different sets of particles, denoted here with subscript $i = 1, 2$, each of them composed by identical rough hard spheres that collide inelastically. Each set is characterized by the following set of mechanical properties: ${m_i, \sigma_i,\kappa_i}$, where $m_i, \sigma_i$ are the particle mass and diameter respectively and $\kappa_i$ is a dimensionless parameter characterizing the moments of inertia $I_i$ of the particles, i.e., $\kappa_i\equiv 4I_i/(m_i\sigma_i^2)$. The velocities of the points of the spheres which are in contact during the collision are

\begin{align}
\centering
\mathbf{w}_i&=\mathbf{v}_i-\frac{\sigma_i}{2}\boldsymbol{\hat\sigma}\times\boldsymbol{\omega}_i, \nonumber \\ 
\mathbf{w}_j&=\mathbf{v}_j+\frac{\sigma_j}{2}\boldsymbol{\hat\sigma}\times\boldsymbol{\omega}_j, \label{relative_velocities_1}
\end{align} where $i, j =1,2$, according to the particle species \cite{K10,SKG10}. Therefore, the relative velocity between colliding particles would be $\mathbf{w}_{ij}\equiv\mathbf{w}_i-\mathbf{w}_j$. In addition to taking into account that linear and angular momentum must be conserved upon collisions, we consider the following relations between pre-collisional and (primed) post-collisional relative velocities

\begin{align}
\centering
  \label{relative_velocities}
\boldsymbol{\hat\sigma}\cdot\mathbf{w}_{ij}'&=-\alpha_{ij}\boldsymbol{\hat\sigma}\cdot\mathbf{w}_{ij}, \nonumber \\
\boldsymbol{\hat\sigma}\times\mathbf{w}_{ij}'&=-\beta_{ij}\boldsymbol{\hat\sigma}\times\mathbf{w}_{ij},   
\end{align} where the former relation stands for the normal component and the latter for the tangential components. Coefficients $\alpha_{ij}, \beta_{ij}$ are the so-called normal and tangential coefficients of restitution. In general, their values depend on the type of binary collision, that in our system can be either gas-gas ($\alpha_{22}, \beta_{22}$) or impurity-gas ($\alpha_{12}, \beta_{12}$), since in the tracer limit, impurity-impurity collisions are statistically rare events and thus they can be neglected \cite{VSG11}. In our collisional model, the coefficients of restitution are independent of the relative velocities of the collision pair.

The kinetic equations for the granular gas and the impurity respectively

\begin{align}
\centering
& \left(\partial_t +\mathbf{v}\cdot\nabla\right) f_2(\mathbf{r}_2,\mathbf{v}_2,\boldsymbol{\omega}_2;t)=J_{22}[\mathbf{r}_2,\mathbf{v}_2,\boldsymbol{\omega}_2;t|f_2,f_2],  \label{KEs2}\\
& \left(\partial_t +\mathbf{v}\cdot\nabla\right) f_1(\mathbf{r}_1,\mathbf{v}_1,\boldsymbol{\omega}_1;t)=J_{12}[\mathbf{r}_1,\mathbf{v}_1,\boldsymbol{\omega}_1;t|f_1,f_2].
\label{KEs1}
\end{align} Here, $J_{ij}$ are the binary collision operators, defined as follows (double primes indicate pre-collisional velocities) \cite{SKG10}

\begin{widetext}
\begin{align}
 J_{ij}[\mathbf{r}_i,\mathbf{v}_i, \boldsymbol{\omega}_i;t|f_i,f_j]\equiv&\sigma_{ij}^2\int\mathrm{d}\boldsymbol{\omega}_j\int\mathrm{d}\mathbf{v}_j\int\mathrm{d}\boldsymbol{\hat\sigma}_i\Theta(\boldsymbol{\hat\sigma}_i\cdot\mathbf{v}_{ij})(\boldsymbol{\hat\sigma}\cdot\mathbf{v}_{ij}) \nonumber \\
& \times\left[(\alpha_{ij}\beta_{ij})^{-2}f_i(\mathbf{r}_i,\mathbf{v}_i'',\boldsymbol{\omega}_i'';t)f_j(\mathbf{r}_j,\mathbf{v}_j'',\boldsymbol{\omega}_j'';t)-f_i(\mathbf{r}_i,\mathbf{v}_i,\boldsymbol{\omega}_i;t)f_j(\mathbf{r}_j,\mathbf{v}_j,\boldsymbol{\omega}_j;t)\right].
\label{coll_integral}
\end{align}
\end{widetext} For solution of the kinetic equations \eqref{KEs2}--\eqref{KEs1}, we may use a multi-temperature Maxwellian form of the velocity distribution function that incorporates non-equipartition of the translational and rotational mode \cite{SKG10,SKS11}:

\begin{equation}
  \label{maxwellian_non_eq}
  f_i(\mathbf{v},\boldsymbol{\omega})=n_i\left(\frac{m_iI_i}{4\pi^2 T_i^\mathrm{tr}T_i^\mathrm{rot}}\right)^{3/2}\exp{\left[m_i\frac{(\mathbf{v}_i-\mathbf{u})^2}{2T_i^\mathrm{tr}}-\frac{I_i\omega^2}{2T^\mathrm{rot}}\right]}.
\end{equation} It is also interesting to recall the definition of the energy production rates 

\begin{equation}
  \xi_{i}^\mathrm{tr}\equiv-\frac{1}{T_i^\mathrm{tr}}\left(\frac{\partial T_i^\mathrm{tr}}{\partial t}\right)_{\mathrm{coll}},  \xi_{i}^\mathrm{rot}\equiv-\frac{1}{T_i^\mathrm{rot}}\left(\frac{\partial T_i^\mathrm{rot}}{\partial t}\right)_{\mathrm{coll}}
\end{equation}
These rates are related to the total cooling rate $\zeta$ as follows 

\begin{equation}
  \label{cooling_rate}
  \zeta=\sum_{i}\frac{n_i}{2nT}\left(T_i^\mathrm{tr}\xi_i^\mathrm{tr} + T_i^\mathrm{rot}\xi_i^\mathrm{rot} \right). 
\end{equation}

%\begin{equation}
%  \label{zeta}
%  \zeta\equiv-\frac{1}{T}\left(\frac{\partial T}{\partial t}\right),
%\end{equation} with $T=\sum \frac{n_i}{2n}(T_i^\mathrm{tr}+T_i^\mathrm{rot})\simeq (T_2^\mathrm{tr}+T_2^\mathrm{rot})/2$ since in the tracer limit $n_1\to 0$.

By introducing this form of the distribution function in \eqref{KEs2}--\eqref{KEs1} and multiplying by $v^2,\omega^2$, we obtain the evolution equations for the translational and rotational temperatures of both species (see \cite{SKG10} for more reference)

\begin{align}
  \label{HCS_eqs}
& \partial_t T=-\zeta T,\\
& \partial_t\frac{T_i^{\mathrm{tr}}}{T} = -\left(\xi_i^{\mathrm{tr}}-\zeta\right)\frac{T_i^{\mathrm{tr}}}{T},\quad \partial_t\frac{T_i^\mathrm{rot}}{T} = -\left(\xi_i^{\mathrm{rot}}-\zeta\right)\frac{T_i^\mathrm{rot}}{T}. \nonumber
\end{align} In the tracer limit, for the impurity we can calculate the functions
 $\xi_1^{\mathrm{tr}}(\frac{T_1^\mathrm{tr}}{T_2^\mathrm{tr}}, \frac{T_1^\mathrm{rot}}{T_2^\mathrm{rot}},\kappa_1, m_1/m_2, \alpha_{12},\beta_{12}, \alpha_{22},\beta_{22})$ and $\xi_1^{\mathrm{rot}}(\frac{T_1^\mathrm{tr}}{T_2^\mathrm{rot}}, \frac{T_1^\mathrm{rot}}{T_2^\mathrm{rot}}, \kappa_1, m_1/m_2, \alpha_{12},\beta_{12}, \alpha_{22},\beta_{22})$ (for the other species can be found in \cite{SKG10}). 
%By using the HCS condition $\xi_1^{\mathrm{tr}}=\xi_1^{\mathrm{tr}}$ and the total cooling rate $\zeta$ \cite{VSK14} we can obtain a closed quartic equation (whose expression is quite large) for one of the temperature ratios $\frac{T_1^\mathrm{tr}}{T_2^\mathrm{tr}}, \frac{T_1^\mathrm{rot}}{T_2^\mathrm{rot}}$.  

% with $T=\sum \frac{n_i}{2n}(T_i^\mathrm{tr}+T_i^\mathrm{rot})\simeq (T_2^\mathrm{tr}+T_2^\mathrm{rot})/2$ since in the tracer limit $n_1\to 0$.

 A normal state with all energy production rates equal ($\xi_1^\mathrm{tr}=\xi_2^\mathrm{tr}=\xi_1^\mathrm{rot}=\xi_2^\mathrm{rot}$) is reached after a short transient. In this state the complete time dependence scales with $T$ (because this state is actually the HCS \cite{H83,SKG10}).
%and thus in scaled time units the system is \textit{stationary}; i.e., $\partial_t=0$. 
We have solved equations \eqref{HCS_eqs} for this state, obtaining the rotational and translational temperatures of both species. 
In the tracer limit the abundant species is decoupled from the impurity and the impurity production rates contributions come only from gas-impurity collisions, which in our case can be done analytically. The temperature ratios $T_1^{\mathrm{tr}}/T_2^{\mathrm{tr}}, T_1^{\mathrm{rot}}/T_2^{\mathrm{rot}}$ are obtained from the physical solution of a quartic equation. For more details on the theoretical procedures refer to a previous work \cite{SKG10}.

We compare our analytical results with an exact numerical solution of \eqref{KEs2}-\eqref{KEs1}, obtained from the Direct Simulation Monte Carlo (DSMC) method \cite{B94}. DSMC was implemented as usual (details can be found elsewhere \cite{VSG11}), taking into account roughness in the collisions. %As it is known, the DSMC method has two main stages (collision and free drift)  of which the second can be omitted in this case since the HCS is completely space-uniform (the same applies for spatially homogeneous states generated by uniform thermostats, see for instance ). 
Additionally, and since in the kinetic equation and hence in the DSMC method the molecular chaos assumption is inherent, comparison with molecular dynamics (MD) simulation results are insightful and for this reason we also implemented this computational method, by using an event driven algorithm for inelastic rough hard spheres \cite{L91}.

We discuss the results in the next section.

% \begin{widetext}
% \beq
% \ffab(\cca,\wwa;\ccb,\wwb)\to \chiab \left(\frac{\ma\mb}{4\pi^2\Tat\Tbt}\right)^{3/2}\exp\left[-\ma\frac{(\cca-\mathbf{u})^2}{2\Tat}-\mb\frac{(\ccb-\mathbf{u})^2}{2\Tbt}\right]
% \far(\wwa)\fbr(\wwb),
% \label{IV.1}
% \eeq
% \end{widetext}
\begin{figure}
\centering
\includegraphics[width=0.42\textwidth,clip]{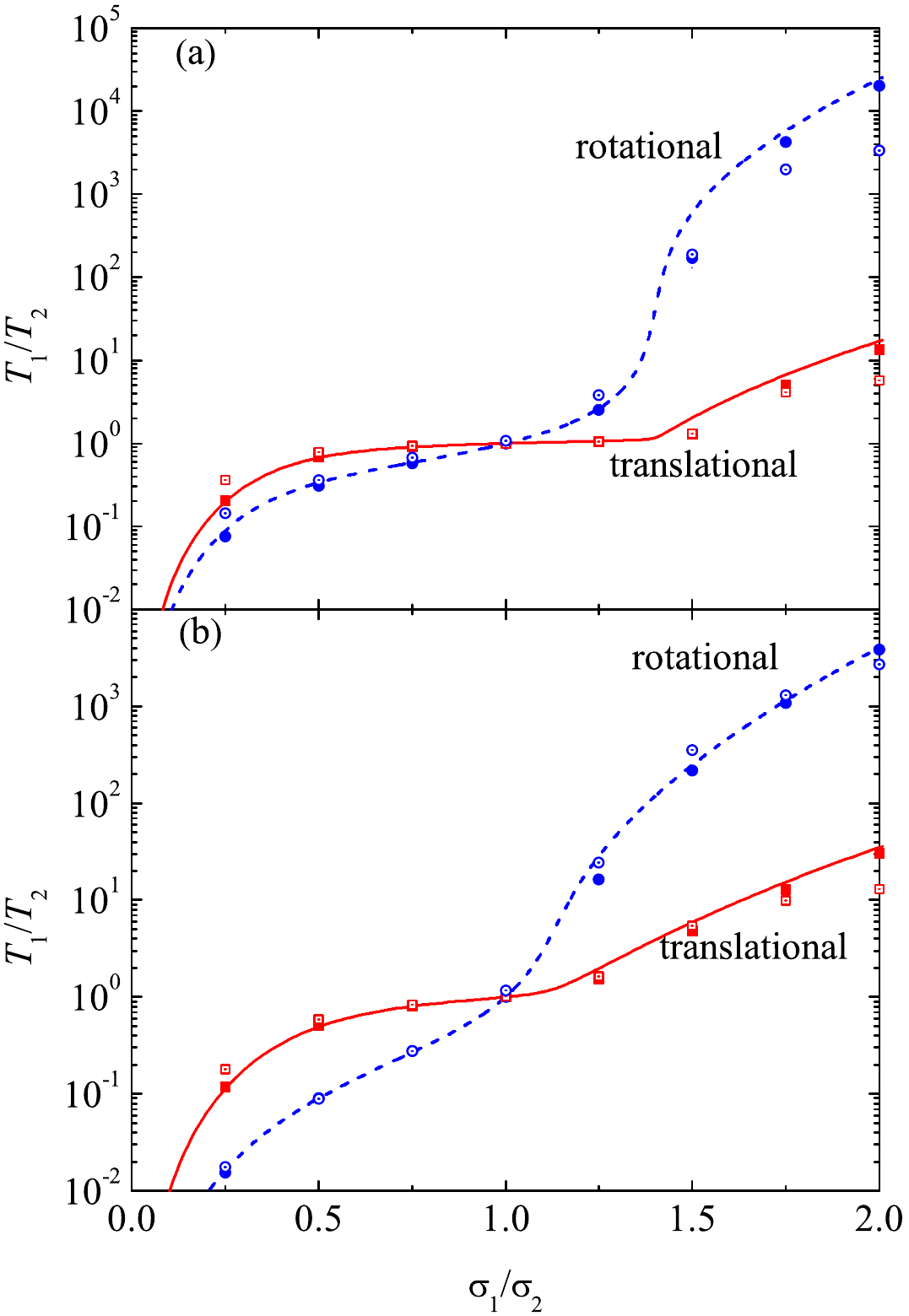}
\caption{Plot of the temperature ratios $T_1^\mathrm{tr}/T_2^\mathrm{tr}$ and $T_1^\mathrm{rot}/T_2^\mathrm{rot}$ vs. the relative size $\sigma_1/\sigma_2$ for relative mass $m_1/m_2 = (\sigma_1/\sigma_2)^3$. (a) $\kappa_1 = \kappa_2=2/5, \alpha_{22} = \alpha_{12} = 0.9$ and $\beta_{22}=\beta_{12}=-0.5$. (b) $\kappa_1 = \kappa_2=2/5, \alpha_{22} = \alpha_{12} = 0.7$ and $\beta_{22}=\beta_{12}=-0.5$. Lines indicate theory results, while solid and open symbols stand for DSMC and MD data, respectively. All MD simulations have a volume fraction $\phi\simeq 0.02$}
\label{sigmaAB}       % Give a unique label
\end{figure}

\section*{Results and discussion}
\label{results}

We present in Fig.\ \ref{sigmaAB} the results of the theory (lines) compared to DSMC (solid symbols) and MD simulations (open symbols) for the case of constant particle mass density ($m_1/m_2=(\sigma_1/\sigma_2)^3$), uniformly distributed inside all particles ($\kappa_1=\kappa_2=2/5$). For simplicity, we chose equal tangential and normal coefficients of restitution for all types of collisions; i.e., $\beta_{22}=\beta_{12}$ and $\alpha_{22}=\alpha_{12}$. As we see in this figure, a good agreement is obtained between theory and both DSMC and MD simulations, for all the range of relative sizes of gas-impurity particles represented here. Even in the context of the multi-temperature Maxwellian approximation used in \eqref{maxwellian_non_eq}, the good agreement could be expected since it has been already been observed for a mono-component granular gas of rough particles that temperature is not very sensitive to the details of the high-energy tails of the velocity distribution function. In a forthcoming work, we will present more cases, extending the results presented here.

\section*{Acknowledgements}
We acknowledge support Grants Nos. FIS2016-76359-P (Spanish Gov.) and  GRU10158 (Junta de Extremadura, partially financed by ERDF funds). Computing facilities from
Extremadura Research Centre for Advanced Technologies (CETA-CIEMAT), funded by the ERDF, are also acknowledged.

%
% BibTeX or Biber users please use (the style is already called in the class, ensure 
% that the "woc.bst" style is in your local directory)
\bibliography{Montpellier17}

\end{document}